\documentstyle[amssymb,aps]{revtex}

\begin{document}
\author{V.Yu. Korda$^1$\thanks{%
Corresponding author, email: kvyu@kipt.kharkov.ua}, A.S. Molev$^1$\thanks{%
Email: mas@kipt.kharkov.ua} and L.P. Korda$^{1,2}$}
\address{$^1$Institute of Electrophysics and Radiation Technologies, National Academy%
\\
of Sciences of Ukraine, 28 Chernyshevsky St., P.O.BOX 8812, UA-61002\\
Kharkov, Ukraine\\
$^2$NSC Kharkov Institute of Physics and Technology, National Academy of\\
Sciences of Ukraine, 1 Akademicheskaya St., UA-61108 Kharkov, Ukraine}
\title{Evolving model-free scattering matrix via evolutionary algorithm: \\
$^{\text{16}}$O-$^{\text{16}}$O elastic scattering at 350 MeV}
\maketitle

\begin{abstract}
We present a new procedure which enables to extract a scattering matrix $%
S\left( l\right) $ as a complex function of angular momentum directly from
the scattering data, without any {\it a priori} model assumptions implied.
The key ingredient of the procedure is the evolutionary algorithm with
diffused mutation which evolves the population of the scattering matrices,
via their smooth deformations, from the primary arbitrary analytical $%
S\left( l\right) $ shapes to the final ones giving high quality fits to the
data. Due to the automatic monitoring of the scattering matrix derivatives,
the final $S\left( l\right) $ shapes are monotonic and do not have any
distortions. For the $^{\text{16}}$O-$^{\text{16}}$O elastic scattering data
at 350 MeV, we show the independence of the final results of the primary $%
S\left( l\right) $ shapes. Contrary to the other approaches, our procedure
provides an excellent fit by the $S\left( l\right) $ shapes which support
the ``rainbow'' interpretation of the data under analysis.

PACS number(s): 24.10.Ht, 25.70.-z, 25.70.Bc
\end{abstract}

\section{Introduction}

$S$-operator is a fundamental quantity of the scattering theory, which
incorporates, by a general assumption, all possible information on any
possible scattering process (including particle creation/destruction). In
the case of an elastic scattering, the diagonal matrix elements of $S$%
-operator in the angular momentum representation can be given in general
form as
\begin{equation}
S\left( l\right) =\eta \left( l\right) e^{2i\varphi \left( l\right) },
\label{S(l)}
\end{equation}
where the $S$-matrix modulus $\eta \left( l\right) $ and the scattering
phase $\varphi \left( l\right) $ are real smooth functions of $l$. The
unitarity of the $S$-matrix for the composite-particle--nucleus scattering
in the presence of nuclear absorption requires that $\eta \left( l\right)
\leq 1$, so we put
\begin{equation}
\eta \left( l\right) =e^{-2\delta _a\left( l\right) },  \label{Mod(S)}
\end{equation}
where the nuclear absorption phase $\delta _a\left( l\right) $ must be a
real smooth positive function of $l$.

Since the colliding nuclei have electric charges, then the scattering phase $%
\varphi \left( l\right) $ can be divided into two parts
\begin{equation}
\varphi \left( l\right) =\delta _r\left( l\right) +\sigma _C\left( l\right) ,
\label{ScatteringPhase}
\end{equation}
where the nuclear refraction phase $\delta _r\left( l\right) $ and the
Coulomb scattering phase $\sigma _C\left( l\right) $ must be real smooth
functions of $l$.

From a general physics viewpoint, the only restrictions we may impose on the
nuclear phases $\delta _{a,r}\left( l\right) $ to be determined are their
finite values at small $l$, total vanishing at sufficiently large $l$ and
smooth behavior in the intermediate region. The most natural and simple
approximation for $\delta _a\left( l\right) $ (or $\eta \left( l\right) $)
and $\delta _r\left( l\right) $ is a monotonically descending (for $\eta
\left( l\right) $, ascending to unity) function which can be easily modelled
with help of, say, the Fermi-step or Gauss functions. For the case of
elastic heavy-ion scattering at intermediate energies ($\gtrsim $20
MeV/nucleon), the $S$-matrix approaches of such a kind (see, e.g., \cite
{McIntyre}, \cite{Kauffmann}, \cite{BerezhnoyPilipenkoMod.Phys.Lett.1995},
\cite{BerezhnoyMolev}) and the optical potential models which yield $S\left(
l\right) $ with such a behavior (see, e.g., \cite{Kobos}, \cite
{BrandanMcVoy.Phys.Rev.C.1991}) have appeared quite successful and argued
for the so-called ``rainbow'' interpretation of the data. However, these
models have not allowed the adequate description of all the features of the
data measured.

At the same time, in many cases the quality of fit can be improved when the
phases $\delta _r\left( l\right) $are modified by the additional surface
terms of different forms (see, e.g., \cite{Hauser}, \cite
{BerezhnoyPilipenkoJ.Phys.G.1985}). Such modifications, in general, make the
$S\left( l\right) $dependence nonmonotonic. Note also the $S$-matrix model
with the additional derivative-like interior term in the absorption phase $%
\delta _a\left( l\right) $ \cite{MolevKordaKorda}. The nonmonotonic behavior
of the described type is also inherent in the scattering matrices found with
help of the optical potentials which have both the standard Saxon-Woods
forms and the ones with the additional surface terms (see, e.g., \cite
{BrandanMcVoy.Phys.Rev.C.1991}, \cite{Austern.Ann.Phys.(N.Y.).15.299.1961.},
\cite{Ershov}). In spite of the nonmonotonicity of $S\left( l\right) $ in
these approaches, the mentioned above ``rainbow'' interpretation of the data
is, nevertheless, preserved.

Further substantial improvement in the quality of fit is achieved with help
of the more flexible $S\left( l\right) $ forms which allow the phases to
behave nonmonotonically for all relevant $l$. Such a nonmonotonic behavior
is provided by extending the standard (monotonic) $S$-matrices with the
series of the pole-like terms (see, e.g., \cite{Allen}) or the proper (say,
spline) basis functions (see, e.g., \cite
{CooperMcEwanMackintosh.Phys.Rev.C45.770.1992.}, \cite
{McEwanCooperMackintosh.Nucl.Phys.A552.401.1993.}, \cite
{CooperMackintosh.Nucl.Phys.A.283.1995.}). The similar behavior is inherent
in the $S$-matrices calculated from the optical potentials which have the
additional derivative-like interior terms or have the more complicated forms
obtained with use of the spline functions or the Fourier-Bessel series (see,
e.g., \cite{Kobos}, \cite{Mermaz}, \cite{Nicoli}, \cite{Ermer}). In spite of
the excellence of the quality of fit provided in such approaches, the
``rainbow'' interpretation of the data appears to be no longer valid, which
rises the problem of finding the physical meaning of the results obtained
this way.

Clearly, all the approaches mentioned above are more or less model-dependent
because the functions used to model the phases $\delta _{a,r}\left( l\right)
$ and the real and imaginary parts of optical potential $V\left( r\right) $
and $W\left( r\right) $ are more or less the properly parametrized
analytical ones. Thus, the search spaces of all possible shapes for the $S$%
-matrix and the optical potential are strongly reduced and, consequently,
the data analyses performed on such spaces can lead to the incorrect
physical interpretation of the data.

That is the reason why it would be highly desirable to have the procedure
which could be able to extract the scattering matrix and/or the optical
potential directly from the experimental data, without introduction of any
bias towards some {\it a priori} ``physically reasonable'' model
assumptions. The very first question this procedure must answer to is
whether the nonmonotonic (or pole-like) structures and any other
distortions, which appear in the $S$-matrix shapes obtained in the most
successful approaches, are really necessary to reproduce the experimental
data studied. This will help us to shed more light on the applicability of
the ``rainbow'' interpretation to the heavy-ion collisions in the wide range
of energies and mass numbers.

\section{Model-free determination of the scattering matrix}

To develope the desired procedure which determines $S\left( l\right) $
directly from the data, $data\longrightarrow S\left( l\right) $, we need to
solve the problem in its most explicit form where each value of $\delta
_{a,r}\left( l\right) $ is treated, generally, as an independent fitting
parameter. This makes the problem parameter space highly dimensional and the
choice of an appropriate search method crucial. Evolutionary (or genetic)
algorithms (EA's) have been many times proven very efficient in dealing with
very difficult physical problems (see, e.g., \cite{Morris}, \cite{Michaelian}%
, \cite{Winkler}, \cite{BerezovskyKordaKlepikov.Phys.Rev.B64.064103.2001.}),
so we have chosen EA as a key element of our procedure. Note that our
algorithm resembles the so-called smooth genetic algorithm proposed in \cite
{Gutowski}.

According to the general ideology of the EA implementation, we deal with the
population of $N$ individuals. Each individual is the $S$-matrix presented
as the pair of the real-valued $l_{\max }$-dimensional vectors $\left(
\delta _a\left( l\right) ,\delta _r\left( l\right) \right) $, $%
l=0,1,...,l_{\max }-1$. The fitness of each individual reflects the quality
of data fitting provided by the individual's $S$-matrix. By using the
mutation operation the algorithm evolves the initial population of the badly
fitted individuals to the final population of the highly fitted ones.

Every iteration of our procedure contains the following steps.

1. Generate the initial population of $N$ individuals. For each individual,
the vectors $\delta _{a,r}\left( l\right) $ are filled with help of any
monotonically descending function of $l$, the first derivative of which has
the only one minimum. To be definite and to test the robustness of the
procedure against various starting conditions, we choose the following five
primary models for $S\left( l\right) $.

a). The 6-parameter model composed of two Fermi functions:
\begin{equation}
2\delta _i\left( l\right) =g_i\,\,\,f\left( l,l_i,d_i\right)
,\,\,\,\,\,\,\,\,\,f\left( l,l_i,d_i\right) =\left[ 1+\exp \left( \frac{l-l_i%
}{d_i}\right) \right] ^{-1},\,\,\,\,\,\,\,\,\,i=a,r.  \label{S(l)_FermyStep}
\end{equation}

b). The 4-parameter model composed of two Gauss functions:
\begin{equation}
2\delta _i\left( l\right) =g_i\,\,\,\exp \left( -\frac{l^2}{d_i^2}\right) .
\label{S(l)_Gauss}
\end{equation}

c). The 5-parameter McIntyre model \cite{McIntyre}:
\begin{equation}
\eta \left( l\right) =f\left( -l,-l_a,d_a\right) ,\,\,\,\,\,\,2\delta
_r\left( l\right) =g_r\,\,\,f\left( l,l_r,d_r\right) .  \label{S(l)_McIntyre}
\end{equation}

d). The 6-parameter phenomenological model \cite{BerezhnoyMolev}:
\begin{eqnarray}
2\delta _i\left( l\right) &=&g_i\left[ \left( 2l+1\right) d_iF\left(
l,l_i,d_i\right) +d_i^2F^2\left( l,l_i,d_i\right) \right]
^{1/2}f^{p_i}\left( l,l_i,d_i\right) ,  \nonumber \\
F\left( l,l_i,d_i\right) &=&-f^{-1}\left( l,l_i,d_i\right) \ln \left[
1-f\left( l,l_i,d_i\right) \right] ,  \nonumber \\
p_a &=&1,\,\,\,p_r=2.  \label{S(l)_Model}
\end{eqnarray}

e). The 6-parameter model composed of two power-type functions:
\begin{equation}
2\delta _i\left( l\right) =\frac{g_i}{l^{\alpha _i}+\beta _i}%
,\,\,\,\,\,\,\,\alpha _i>2.  \label{S(l)_Power}
\end{equation}

The parameters $g_i$, $l_i$, $d_i$, $\alpha _i$, and $\beta _i$, are
positive. They are chosen for each individual and each model function at
random within some intervals which are wide enough to produce substantially
different shapes of the phases. Normally, all the individuals in a given
population are initialized with one and the same function from the set a) -
e).

All the mentioned primary models for $S\left( l\right) $ are ``physically
justified'', except for the case e) which has no physical background.
Nevertheless, we have included this purely mathematical case to see whether
the procedure is able to find ``physically meaningful'' results under such a
tough conditions.

2. Evaluate the fitness of each individual in the population. The fitness
function in our approach consists of two parts. The first one is associated
with the quality of the shapes of $\delta _{a,r}\left( l\right) $ while the
second one accounts for the quality of the fitting of the experimental data.

The requirements which the shapes of $\delta _{a,r}\left( l\right) $ must
meet in our approach are as follows.

i). The functions $\delta _{a,r}\left( l\right) $ must be descending.

ii). The first derivatives of $\delta _{a,r}\left( l\right) $ must have only
one minimum and no maxima.

iii). The second derivatives of $\delta _{a,r}\left( l\right) $ must not
have more than one minimum and one maximum.

iv). The third derivative of $\delta _r\left( l\right) $ must not have more
than one minimum and one maximum.

v). The logarithmic derivative of $\delta _r\left( l\right) $ must be
descending.

The requirements i) - iii) ensure the absence of any distortions of the
phase shapes, at least, up to the derivatives of the second order. The
condition iv) is added because we want the deflection function $\Theta
\left( l\right) \equiv 2d\varphi \left( l\right) /dl$ to have no shape
distortions up to the same order of its derivatives. The condition v)
provides for the permanent decrease of $\delta _r\left( l\right) $ with the
increase of $l$. The requirements i) - iv) are crucial for the shapes of $%
\delta _{a,r}\left( l\right) $. Thus the penalties imposed upon the
individual in the case of the violation of these requirements are fatal. The
condition v) is not so strong and introduces only the ultimate bias towards
the desired tail of $\delta _r\left( l\right) $.

The quality of the fit of the calculated differential cross section to the
experimentally measured one is assessed via the standard $\chi ^2$ magnitude
per data point. The calculations are made by using the expansion of the
scattering amplitude into a series of Legendre polynomials.. The elastic
scattering differential cross section is equal to the squared modulus of
this amplitude.

It is often claimed that the amount of the large scattering angle data is
insufficient to determine the scattering matrix and/or the optical potential
in a unique way. Thus, we add several additional pseudo data points after
the last actual ones, which follow the tendency of the cross section
behavior (cf., e.g., \cite{McEwanCooperMackintosh.Nucl.Phys.A552.401.1993.}%
). Of course, this prescription can not be universal and must be used with
care in the context of the data under study. The incorporation of the
invented data points to the $\chi ^2$ criterium can appear misleading for
the fitting procedure, therefore, we use the penalty-free corridor around
those points and apply the prescription only after the fitting to the actual
data set has been accomplished.

3. Let each individual in the population produce $M$ offsprings. The
replication is performed according to the transformation:
\begin{equation}
\log \left( \delta _i^{\prime }\left( l\right) \right) =\log \left( \delta
_i\left( l\right) \right) +A_i\,\,\,N_i\left( 0,1\right) \,\,\,D\left(
l,\,\,l_{m,i},\,\,d_{m,i}\right) ,\,\,\,\,\,\,\,\,\,i=a,r,
\label{ReplicationTransformation}
\end{equation}
where $\delta _i\left( l\right) $ and $\delta _i^{\prime }\left( l\right) $
are the parent's and offspring's $S$-matrix phases, respectively, $A_i>0$ is
the mutation amplitude, $N_i\left( 0,1\right) $ denotes a normally
distributed one-dimensional random number with mean zero and one standard
deviation, $l_{m,i}$ stands for the mutation point chosen randomly in the
interval $0\leq l_{m,i}\leq l_{\max }-1$, and $d_{m,i}>0$ is the value
characterizing the diffuseness of the mutation point. The diffusing function
$D\left( l,\,\,l_{m,i},\,\,d_{m,i}\right) $ must be of the bell-like shape
with the only maximum at $l=l_{m,i}$ and the fall-off tail around this
point. To be definite and to ensure the proper localization of the
consequences of the mutation we choose the diffusing function in the form:
\begin{equation}
D\left( l,\,\,l_{m,i},\,\,d_{m,i}\right) =\exp \left[ -\frac{\left(
l-l_{m,i}\right) ^2}{d_{m,i}^2}\right] .  \label{DiffusingFunction}
\end{equation}

The mutation amplitude $A_i$ and the mutation diffuseness $d_{m,i}$ are the
quantities automatically tuned within some intervals. The limits of these
intervals, having the extremely large values at the beginning of the
procedure, are smoothly decreased in the course of the run, and acquire the
small values at the end. Such a schedule provides for both the removal of
the features of the primary parameterizations a) - e) from the individual's $%
S\left( l\right) $ and the fine tuning of the details of $S\left( l\right) $.

4. Evaluate the fitness values of all offsprings. Sort the offsprings in a
descending order due to their fitnesses. Select $N$ best offsprings to form
the new population.

5. Go to step 3 or stop if the best fitness in the population is
sufficiently high (the $\chi ^2$ value is small enough).

Evolutionary algorithms are, generally, the global optimization technique
which, however, can not guarantee that the optimum found is the global one.
Therefore, it is necessary to run the procedure several times. Besides,
there is no way to know in advance what will be the minimum value of the $%
\chi ^2$ magnitude. Thus, it is instructive to monitor the dynamics of the
best, worst and mean fitness values and the rms deviation from the mean
fitness in the population during those several runs of the procedure. Such
monitoring usually helps to localize the region of the potentially lowest $%
\chi ^2$ values.

\section{Scattering matrix for the $^{\text{16}}$O-$^{\text{16}}$O elastic
scattering at 350 MeV}

We have applied our technique to analyze the well known test case of the $^{%
\text{16}}$O-$^{\text{16}}$O elastic scattering at 350 MeV, for which the
approaches giving very good quality of fit predict the existence of the
nonmonotonic structures in the $S$-matrix (see, e.g., \cite{Allen}, \cite
{CooperMcEwanMackintosh.Phys.Rev.C45.770.1992.}).

In our calculations, bearing in mind that the collision energy is
sufficiently high, we let $\sigma _C\left( l\right) $ in Eq. (\ref
{ScatteringPhase}) to be the quasiclassical phase of the point-charge
scattering by the uniformly charged sphere (see, e.g., \cite
{BerezhnoyPilipenkoMod.Phys.Lett.1995}) having the radius $R_C=0.95\times
2\times 16^{1/3}$ \cite{BrandanSatchler}. The calculated differential cross
sections have been symmetrized for the scattering of identical nuclei.
Besides, the experimental errors are assumed to be equally weighted (10\%
error bars).

Figures 1-5 show the results of our calculations with the primary models a)
- e) for $S\left( l\right) $, respectively. The $\chi ^2$ values for our
fits to the data are $2.4-2.5$. For each initial case, the results of five
different runs of the procedure are presented to display the error bands
within each of the primary $S\left( l\right) $ models. Figure 6 compiles
five best results from Figs. 1-5 to illuminate their sensitivity to the
details of the particular primary $S\left( l\right) $ model. Figure 7
demonstrates the consequences of the consideration of the invented data
points in the region of large scattering angles.

\section{Discussion}

The evolutionary procedure of determining the scattering matrix, presented
in this publication, is aimed to search for the globally optimal solution.
But, being aware of the complexity of the problem under study and the fact
that the actual number of fitting parameters (twice the number of angular
momenta which is $l_{\max }=$120 in our test case) is substantially greater
the actual number of data points (which is equal to 105 in our test case),
we do not expect to achieve it. Therefore, we consider the obtained results
(Figs. 1-7) as very promising.

First of all, we see that within every model used for the primary $S\left(
l\right) $ dependence, regardless the variety of their shapes, the moduli $%
\eta \left( l\right) $ and the nuclear refraction phases $\delta _r\left(
l\right) $, as well as the total deflection functions $\Theta \left(
l\right) $, obtained in different runs of the procedure, go close to each
other (Figs. 1-5). The differences between them can sometimes be seen only
in the enlarged or even logarithmic scale. The same observation can be made
if one analyzes the compilation of the best results (Fig. 6), which points
on their independence upon the initial conditions. At the same time, the
nuclear phases $\delta _r\left( l\right) $ deviate from each other in the
region of large angular momenta. There the scattering matrix moduli $\eta
\left( l\right) $ are very close to unity, which makes the contributions of
the partial waves with these values of $l$ to the scattering amplitude
vanishingly small. To introduce the corresponding bias into the searching
procedure, we probably need more precise experimental information in the
region of small scattering angles. Nevertheless, we are able to conclude
that, under the requirements i) - v) imposed upon the phases $\delta
_{a,r}\left( l\right) $, we have managed to localize the region of the
scattering matrix shapes which give the lowest values to the $\chi ^2$
magnitude.

It is important to emphasize the remarkable fact that the application of the
power-type function e) [Eq. (\ref{S(l)_Power})] as the primary $S\left(
l\right) $ parameterization, which has no proper physical meaning, does not
produce any difficulties for our procedure to find the physically meaningful
scattering matrix (Fig. 5). From the formally mathematical viewpoint, the
iterative application of the replication transformation (\ref
{ReplicationTransformation}) to the phases $\delta _{a,r}\left( l\right) $
is equivalent to the addition to the primary phases of the ultimately
``infinite'' sum of the diffusing functions (\ref{DiffusingFunction}) with
various parameters and weights. Due to the special schedule of choosing and
tuning the latter, the phase shapes are transformed almost adiabatically
across the run of the procedure. As the result, the phases evolve to the
equilibrium, with respect to the fitness, shapes which are free from any
recollections about the particular models for the primary $S\left( l\right) $
and the diffusing function. That is why, we believe, our procedure is
actually a model-free one.

Somewhat surprising seems the observation that the incorporation of the
additional pseudo data points in the fitness function, which really forces
the cross section to behave as desired, produces no noticeable corrections
to the scattering matrix (Fig. 7) in the whole range of $l$. This is against
the conventional way of thinking but can be just the feature of that
particular data set under study.

From the physics viewpoint, our results support the ``rainbow''
interpretation of the given data: the maximum in the differential cross
section observed near 50$^{\circ }$ is identified as the primary nuclear
rainbow. The nuclear rainbow angle which corresponds to the minimum of the
deflection function $\Theta \left( l\right) $ acquires the values $\theta
_r=61-64^{\circ }$. At this point, one might ask whether it is possible to
improve the quality of fit, bearing in mind the number of fitting
parameters. In fact, the answer is positive. If, from the very beginning, we
abandon all the requirements i) - v) imposed on the shapes of $\delta
_{a,r}\left( l\right) $, then the procedure becomes able to find the results
with $\chi ^2\approx 0.5-0.6$. But the $S$-matrices for these cases are
nonmonotonic and substantially different from run to run, belonging to
different local optima. The other way to search for the better quality of
fit could be found in testing the stability of the monotonic shapes of the $%
S $-matrices obtained in our study against the nonmonotonic transformations (%
\ref{ReplicationTransformation}). Then it seems more probable to find the
results which belong to the same local optimum or the nearest ones.
Following this way, if and only if the substantial improvement in the
quality of fit is accompanied by the repeated observations of the same
equilibrium nonmonotonic structures in $S\left( l\right) $, then the
appearance of these structures should be admitted as necessary and the
search for their physical interpretation urgent.

The proposed recipe can also be useful in analyzing some important cases
where the presence of nonmonotonic (or even nonsmooth) structures in the $S$%
-matrix seems to be justified. Namely, these are the cases where, for
instance, the behavior of $S\left( l\right) $ is resonance-dominated (see,
e.g., \cite{BrandanMcVoy.Phys.Rev.C.1991}, \cite
{McVoy.Phys.Rev.C.3.1104.1971.}), or it is important to account for the
dynamic effects (parity dependence of the interactions between nuclei,
elastic transfer, see, e.g., \cite{Frahn.Nucl.Phys.A337.324.1980.}, \cite
{FrahnHusseinCantoDonangelo.Nucl.Phys.A369.166.1981.}), or the interference
effects condition the nonmonotonic scattering matrices (see, e.g., \cite
{Austern.Ann.Phys.(N.Y.).15.299.1961.}). In order to confirm the existence
of the discussed equilibrium structures in the $S$-matrix, we need to
extract them directly from the respective experimental data, using, for
instance, our approach with the requirements i) - v) switched off either
from the very beginning or after the monotonic shape of the $S$-matrix is
obtained. If we succeed, then we must admit that the requirements i) - v)
cannot apply to all cases.

The evolutionary procedure presented in this publication has been initially
devised to determine the scattering matrix in the angular momentum
representation. Obviously, the similar approach can be used to develop the
evolutionary procedure for the determination of radial dependence of a
complex optical potential. With help of this procedure, the optical
potential can be extracted directly from the experimental data: $%
data\longrightarrow V\left( r\right) $. Moreover, using the similar
procedure, the scattering matrix produced by the optical potential can be
fitted to the scattering matrix extracted directly from the data: $%
data\longrightarrow S\left( l\right) $ $\longrightarrow V\left( r\right) $.
This means that the optical potential found this way will correspond to the
scattering matrix extracted immediately from the data. Having unified these
three search procedures $data\longrightarrow S\left( l\right) $, $%
data\longrightarrow V\left( r\right) $, and $data\longrightarrow S\left(
l\right) $ $\longrightarrow V\left( r\right) $ into the one, we obtain a
powerful tool for the deep theoretical investigation of the heavy-ion
collisions at intermediate energies.

\vskip 0.8cm

\begin{center}
{\bf Acknowledgments}
\end{center}

This work was supported in part by The State Fund of Fundamental Research of
Ukraine Grant No.02.07/372.

\begin{center}
\newpage {\bf Figure Captions.}
\end{center}

FIG. 1. Five scattering matrices for the $^{\text{16}}$O-$^{\text{16}}$O
elastic scattering at 350 MeV, calculated by our procedure with the primary
model a) for $S\left( l\right) $ [Eq. (\ref{S(l)_FermyStep})]. (a)
Scattering matrix moduli $\eta \left( l\right) $. The inset shows the region
of small momenta in the logarithmic scale. (b) Nuclear phases $\delta
_r\left( l\right) $. The inset shows the region of small momenta in the
enlarged scale. (c) The same as (b) but in the logarithmic scale. (d)
Deflection functions $\Theta \left( l\right) $. The inset shows the vicinity
of $\Theta \left( l\right) $ minima in the enlarged scale. Solid curves
correspond to the best quality of fit to the data $\chi ^2=2.4$.

FIG. 2. The same as FIG. 1 but with the primary model b) for $S\left(
l\right) $ [Eq. (\ref{S(l)_Gauss})].

FIG. 3. The same as FIG. 1 but with the primary model c) for $S\left(
l\right) $ [Eq. (\ref{S(l)_McIntyre})].

FIG. 4. The same as FIG. 1 but with the primary model d) for $S\left(
l\right) $ [Eq. (\ref{S(l)_Model})].

FIG. 5. The same as FIG. 1 but with the primary model e) for $S\left(
l\right) $ [Eq. (\ref{S(l)_Power})].

FIG. 6. Five best results from Figs. 1-5. Notations are the same as in FIG.
1.

FIG. 7. Two scattering matrices and differential cross sections for the $^{%
\text{16}}$O-$^{\text{16}}$O elastic scattering at 350 MeV, calculated by
our procedure with the primary model b) for $S\left( l\right) $ [Eq. (\ref
{S(l)_Gauss})]. Solid (dashed) curves are the results of calculations with
the invented data points taken (not taken) into account in the region of
large scattering angles. (a) Scattering matrix moduli $\eta \left( l\right) $
and (b) nuclear phases $\delta _r\left( l\right) $ in the region of small
momenta. (c) $\delta _r\left( l\right) $ in the region of large momenta. (d)
Deflection functions $\Theta \left( l\right) $ in the vicinity of the
minima. (e) The differential cross sections (ratio to Rutherford).
Experimental data are taken from Refs. \cite{Stilliaris} and \cite
{BrandanSatchlerPhysLett}. Solid curves presenting $S\left( l\right) $
correspond to the same ones shown on FIG. 2.

\vskip 0.8cm

{\bf Note to the Editor.} Please, place each figure on a single journal page.


\begin{references}
\bibitem{McIntyre}  J.A. McIntyre, K.H. Wang and L.C. Becker, Phys. Rev.
{\bf 117}, 1337 (1960).

\bibitem{Kauffmann}  S.K. Kauffmann, Z. Phys. A {\bf 282}, 163 (1977).

\bibitem{BerezhnoyPilipenkoMod.Phys.Lett.1995}  Yu.A. Berezhnoy and V.V.
Pilipenko, Mod. Phys. Lett. A {\bf 10}, 2305 (1995).

\bibitem{BerezhnoyMolev}  Yu.A. Berezhnoy and A.S. Molev, Int. J. Mod. Phys.
E {\bf 12}, 827 (2003).

\bibitem{Kobos}  A.M.Kobos, M.E. Brandan, and G.R. Satchler, Nucl. Phys.
{\bf A487}, 457 (1988).

\bibitem{BrandanMcVoy.Phys.Rev.C.1991}  M.E. Brandan and K.W. McVoy, Phys.
Rev. C {\bf 43}, 1140 (1991).

\bibitem{Hauser}  G. Hauser, R. L\"{o}hken, H. Rebel, G. Schatz, G.W.
Schweimer, and J. Specht, Nucl. Phys. {\bf A128}, 81 (1969).

\bibitem{BerezhnoyPilipenkoJ.Phys.G.1985}  Yu.A. Berezhnoy and V.V.
Pilipenko, J. Phys. G. {\bf 11}, 1161 (1985).

\bibitem{MolevKordaKorda}  A.S. Molev, V.Yu. Korda, and L.P. Korda, Izv.
Ross. Acad. Nauk, Ser. Fiz. {\bf 68}, 205 (2004).

\bibitem{Austern.Ann.Phys.(N.Y.).15.299.1961.}  N. Austern, Ann. Phys.
(N.Y.) {\bf 15}, 299 (1961).

\bibitem{Ershov}  S.N. Ershov, F.A. Gareev, R.S. Kurmanov, E.F. Svinareva,
S.G. Kazacha, A.S. Dem'yanova, A.A. Ogloblin, S.A. Goncharov, J.S. Vaagen,
and J.M. Bang, Phys. Lett. B {\bf 227}, 315 (1989).

\bibitem{Allen}  L.J. Allen, L. Berge, C. Steward, K. Amos, H. Fiedeldey, H.
Leeb, R. Lipperheide, and P. Fr\"{o}brich, Phys. Lett. B {\bf 298}, 36
(1993).

\bibitem{CooperMcEwanMackintosh.Phys.Rev.C45.770.1992.}  S.G. Cooper, M.A.
McEwan, and R.S. Mackintosh, Phys. Rev. C {\bf 45}, 770 (1992).

\bibitem{McEwanCooperMackintosh.Nucl.Phys.A552.401.1993.}  M.A. McEwan, S.G.
Cooper, and R.S. Mackintosh, Nucl. Phys. {\bf A552}, 401 (1993).

\bibitem{CooperMackintosh.Nucl.Phys.A.283.1995.}  S.G. Cooper and R.S.
Mackintosh, Nucl. Phys. {\bf A582}, 283 (1995).

\bibitem{Mermaz}  M.C. Mermaz, Phys. Rev. C {\bf 47}, 2213 (1993).

\bibitem{Nicoli}  M.P. Nicoli, F. Haas, R.M. Freeman, S. Szilner, Z. Basrak,
A. Morsad, G.R. Satchler, and M.E. Brandan, Phys. Rev. C {\bf 61}, 034609
(2000).

\bibitem{Ermer}  M. Ermer, H. Clement, G. Frank, P. Grabmayr, N. Heberle,
and G.J. Wagner, Phys. Lett. B {\bf 224}, 40 (1989).

\bibitem{Morris}  J.R. Morris, D.M. Deaven, and K.M. Ho, Phys. Rev. B {\bf 53%
}, R1740 (1996).

\bibitem{Michaelian}  K. Michaelian, Revista Mexicana de Fisica {\bf 42 }%
(suppl.1), 203 (1996).

\bibitem{Winkler}  C. Winkler and H.M. Hofmann, Phys. Rev. C {\bf 55}, 684
(1997).

\bibitem{BerezovskyKordaKlepikov.Phys.Rev.B64.064103.2001.}  S.V.
Berezovsky, V.Yu. Korda and, V.F. Klepikov, Phys. Rev. B {\bf 64}, 064103
(2001).

\bibitem{Gutowski}  M.W. Gutowski, J. Phys. A {\bf 27}, 7893 (1994).

\bibitem{BrandanSatchler}  M.E. Brandan and G.R. Satchler, Nucl. Phys. {\bf %
A487}, 477 (1988).

\bibitem{Stilliaris}  E. Stilliaris, H.G. Bohlen, P. Fr\"{o}brich, B.
Gebauer, D. Kolbert, W. von Oertzen, M. Wilpert, and Th. Wilpert, Phys.
Lett. B {\bf 223}, 291 (1989).

\bibitem{BrandanSatchlerPhysLett}  M.E. Brandan and G.R. Satchler, Phys.
Lett. B {\bf 256}, 311 (1991).

\bibitem{McVoy.Phys.Rev.C.3.1104.1971.}  K.W. McVoy, Phys. Rev. C {\bf 3},
1104 (1971).

\bibitem{Frahn.Nucl.Phys.A337.324.1980.}  W.E. Frahn, Nucl. Phys. {\bf A337}%
, 324 (1980).

\bibitem{FrahnHusseinCantoDonangelo.Nucl.Phys.A369.166.1981.}  W.E. Frahn,
M.S. Hussein, L.F. Canto, and R. Donangelo, Nucl. Phys. {\bf A369}, 166
(1981).
\end{references}
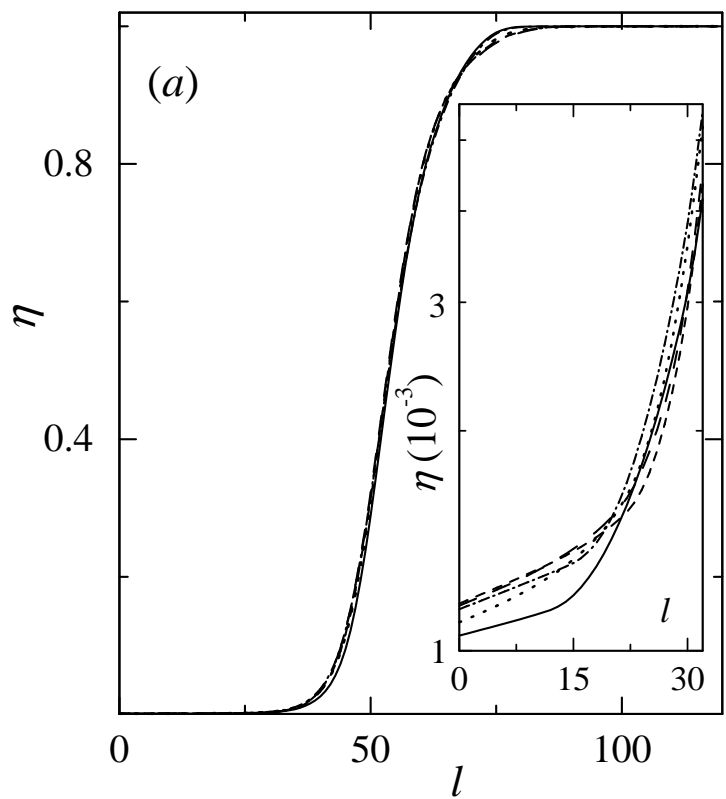
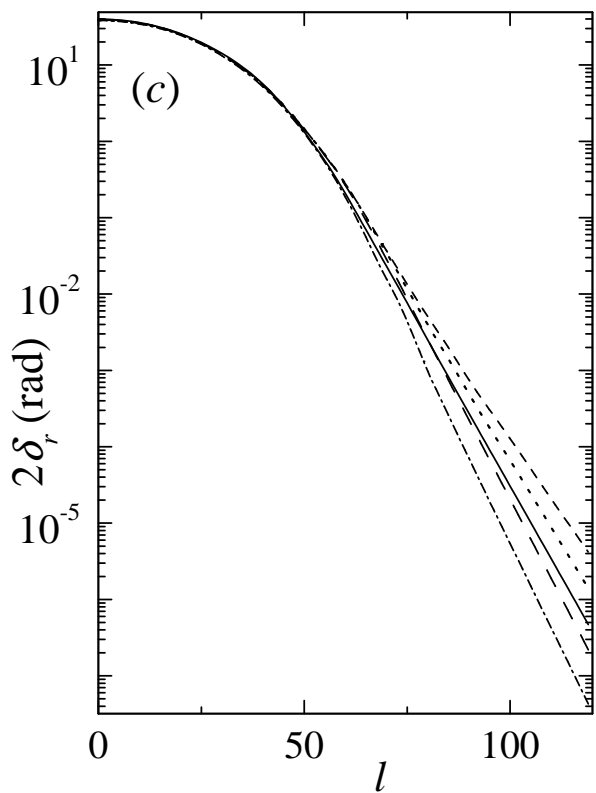
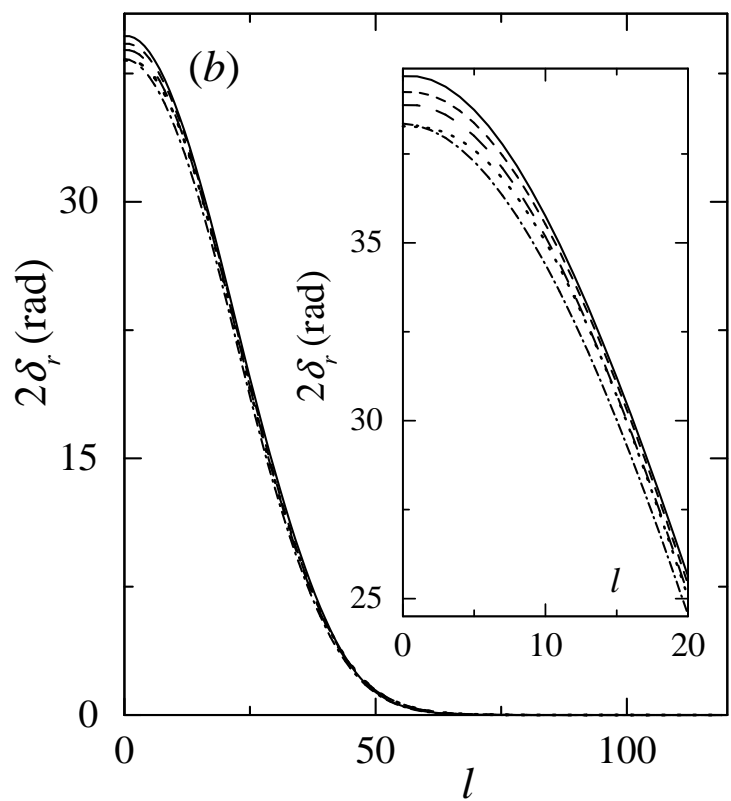
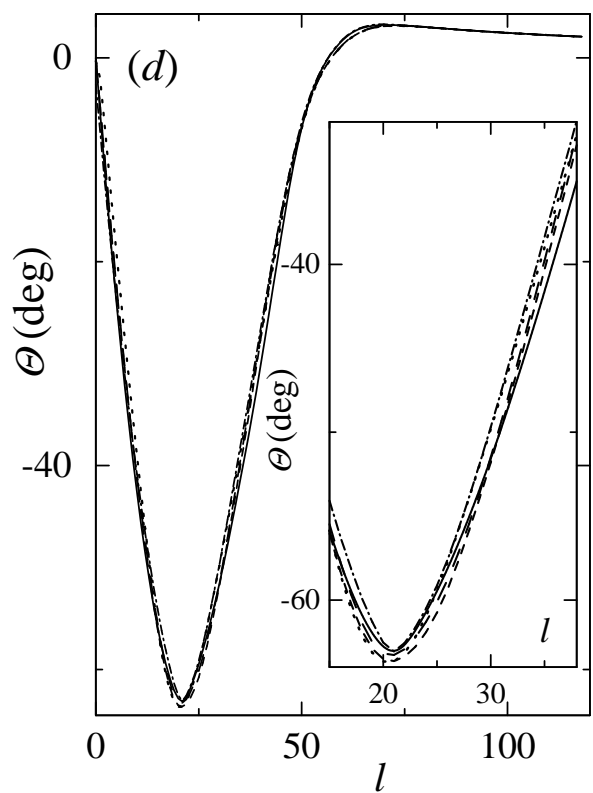
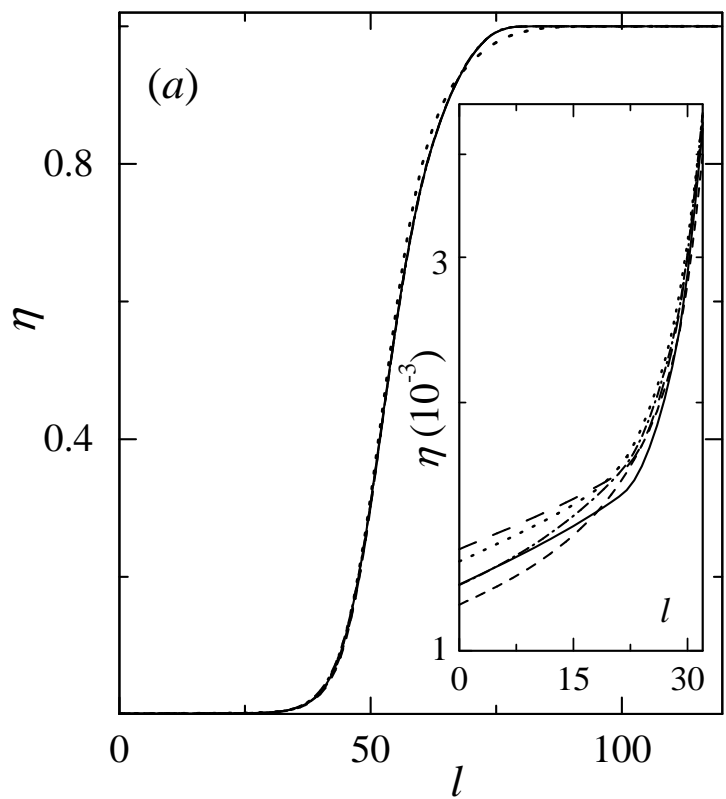
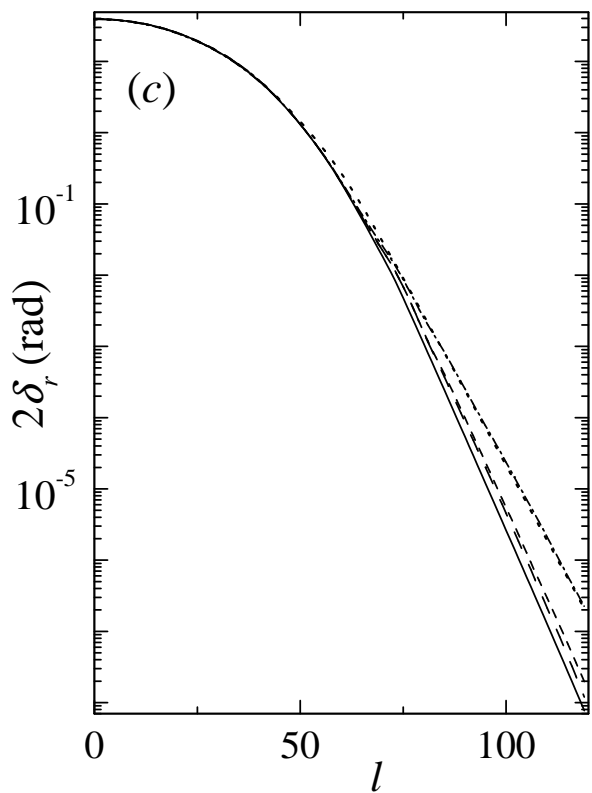
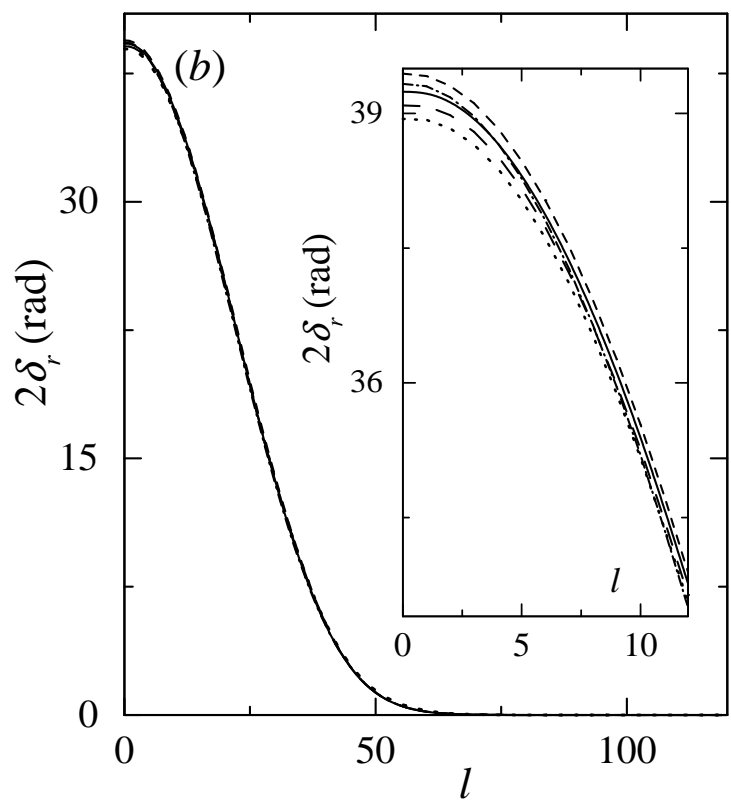
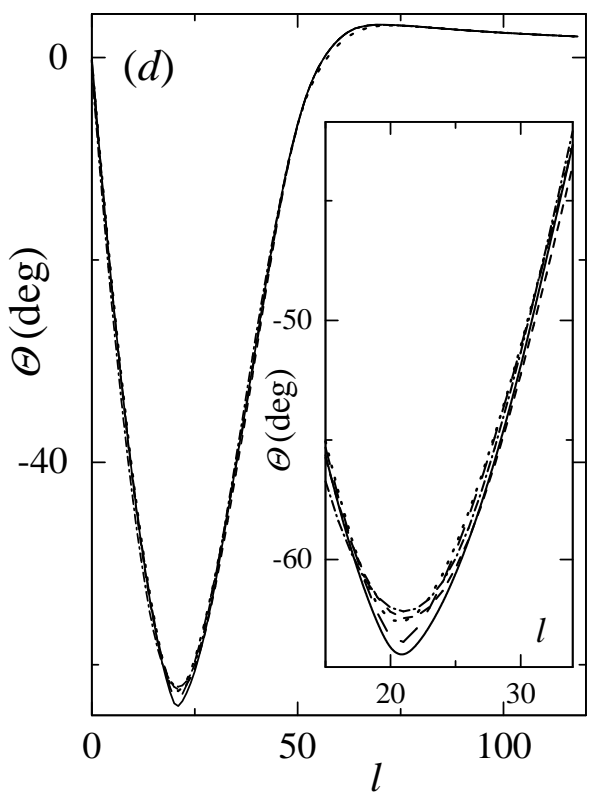
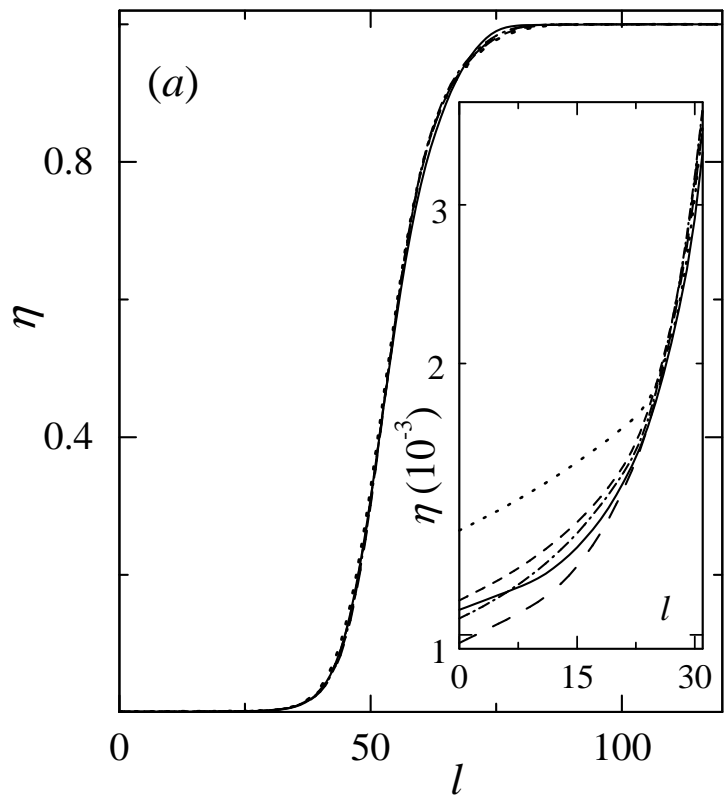
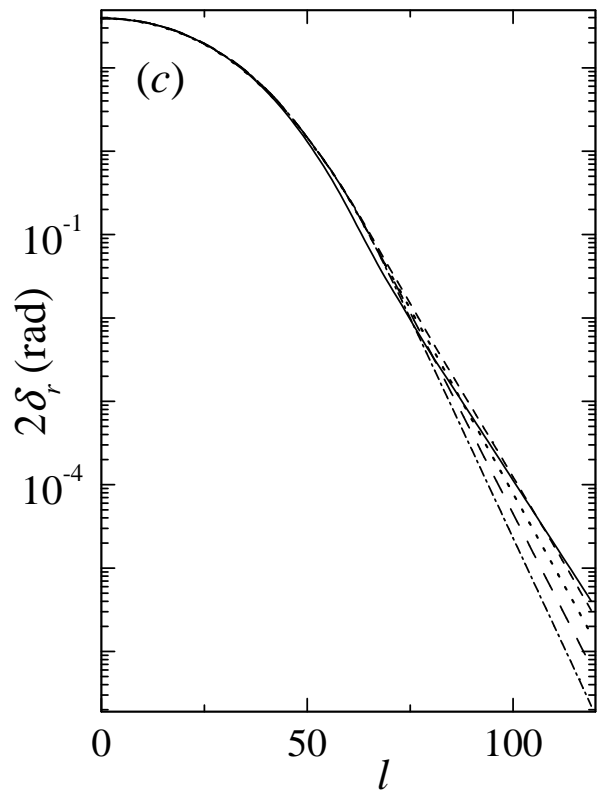
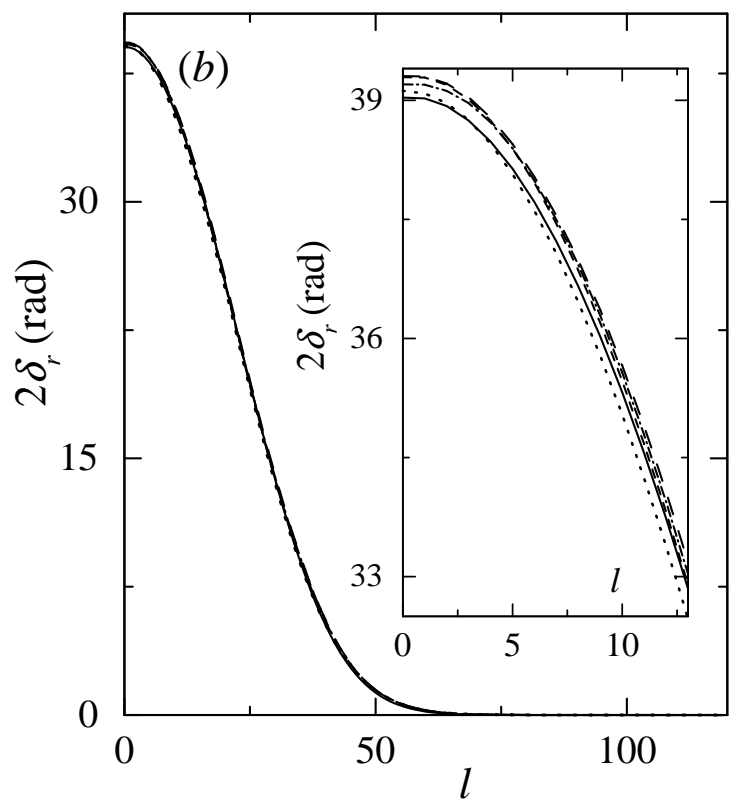
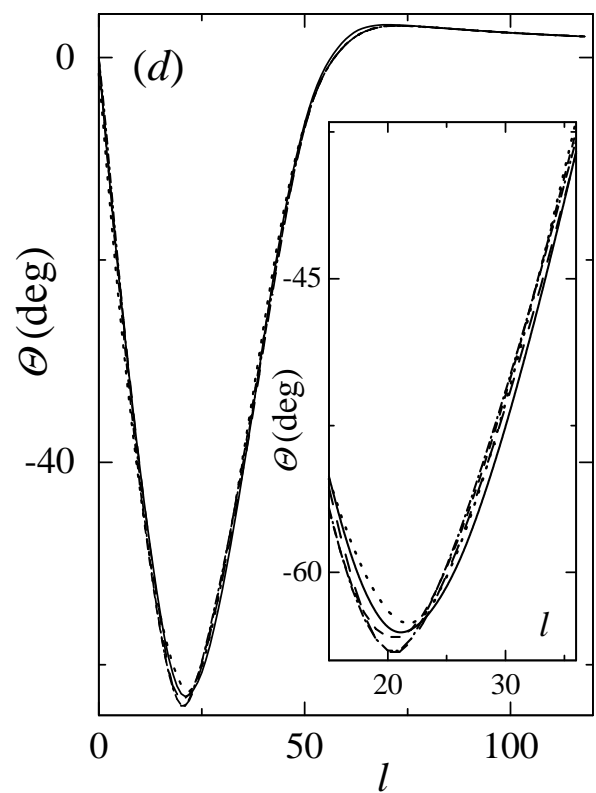
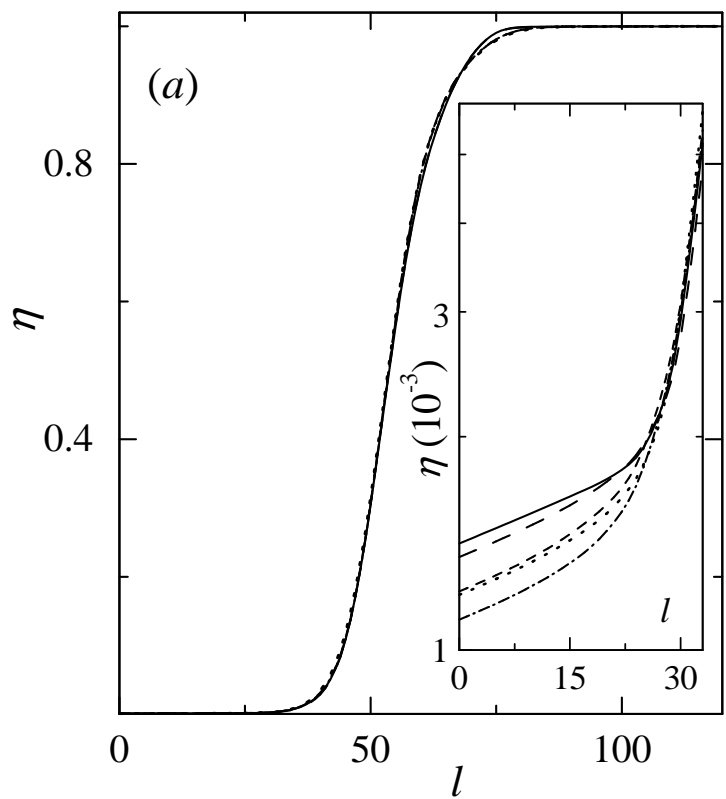
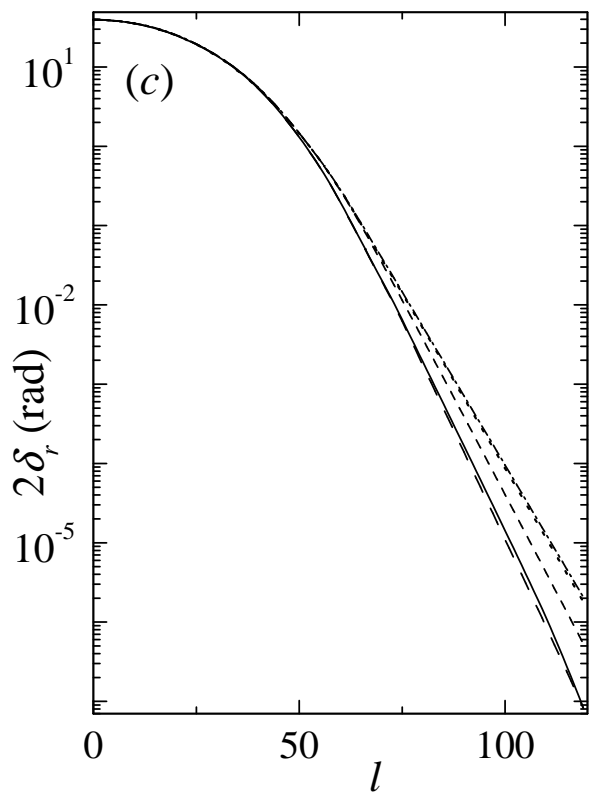
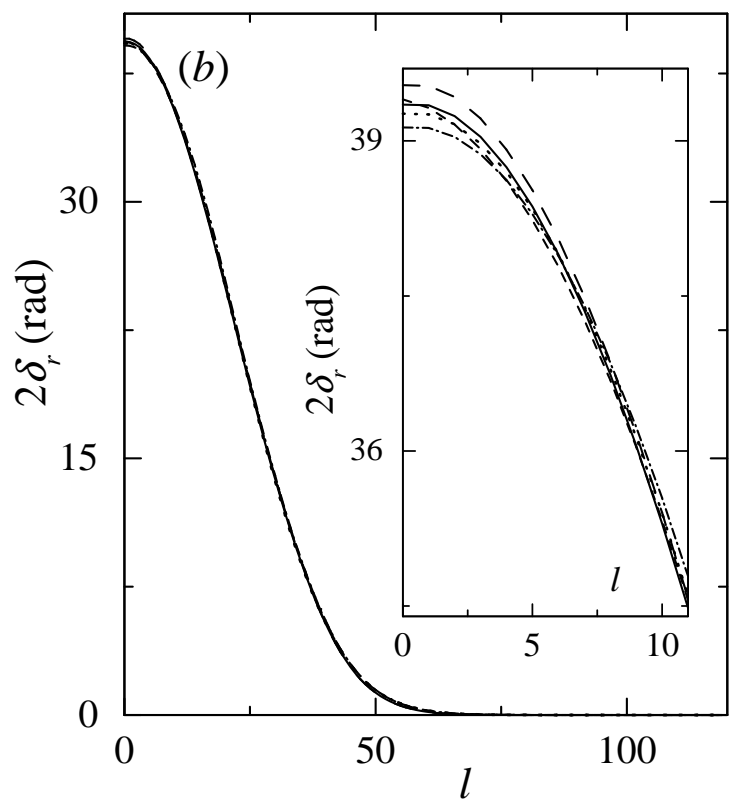
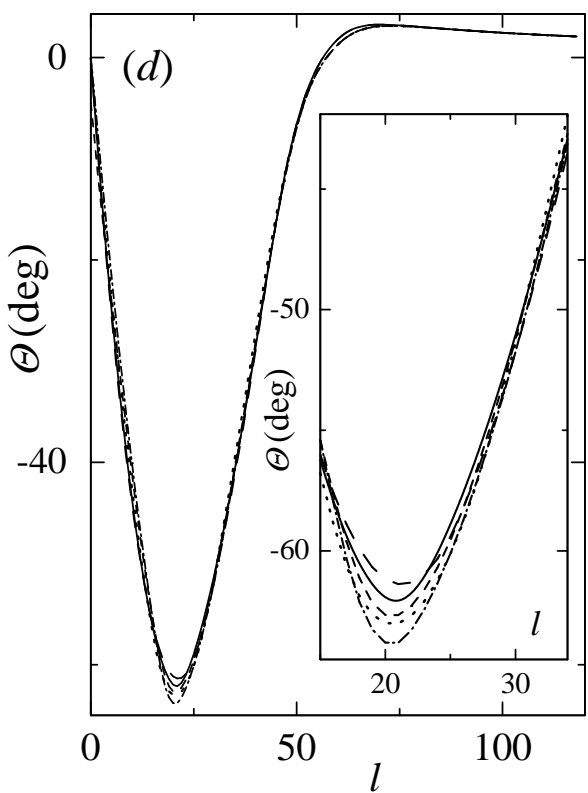
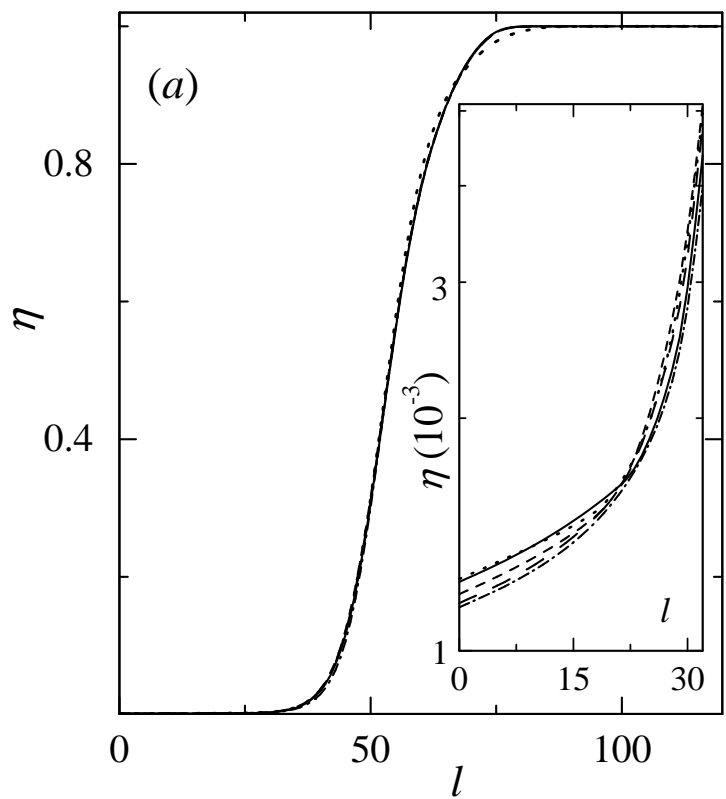
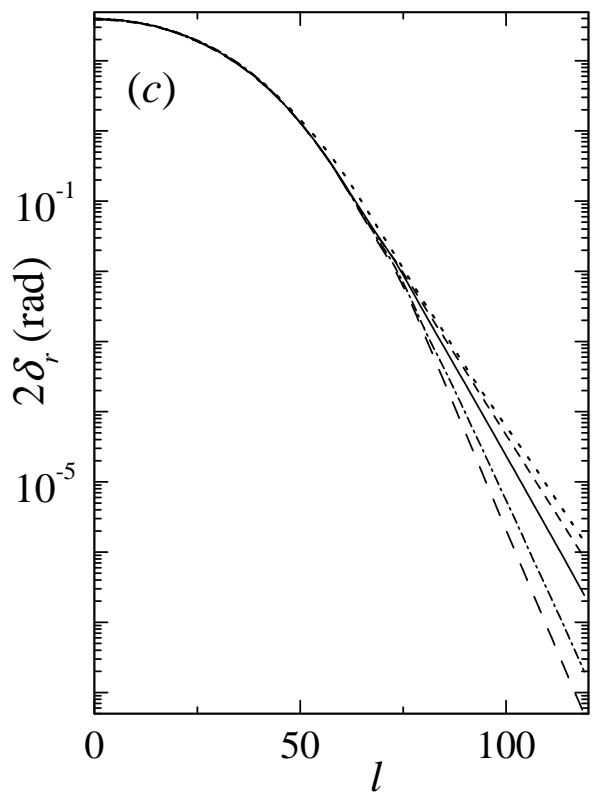
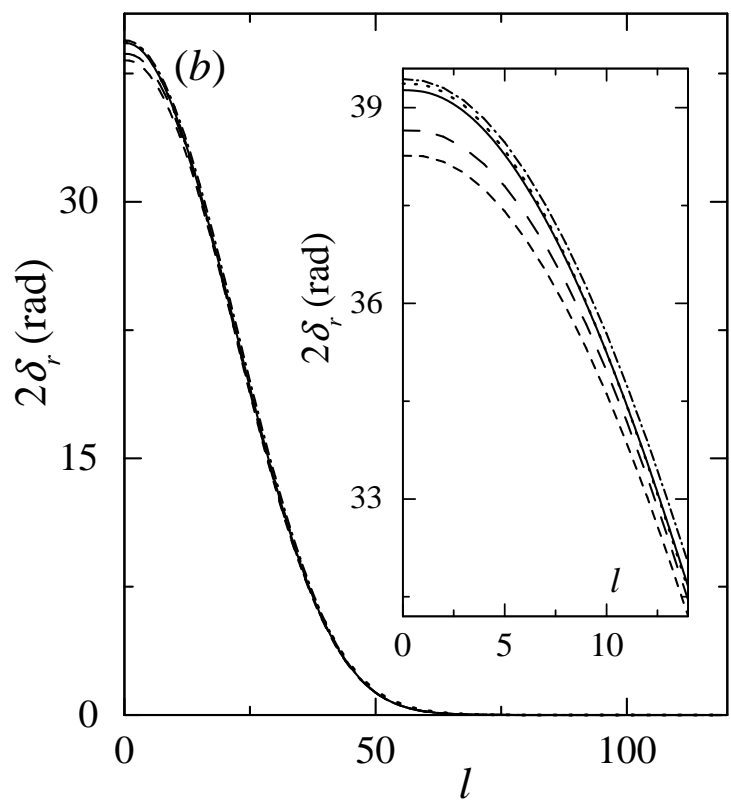
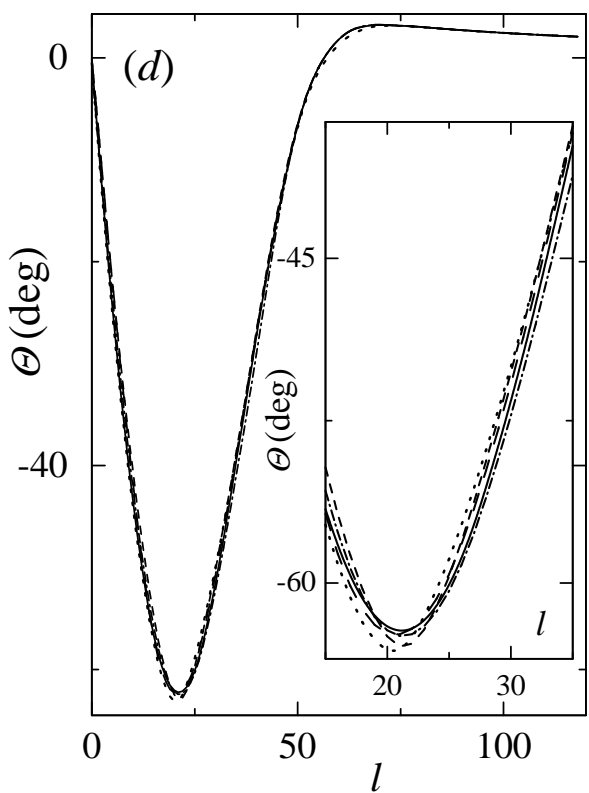
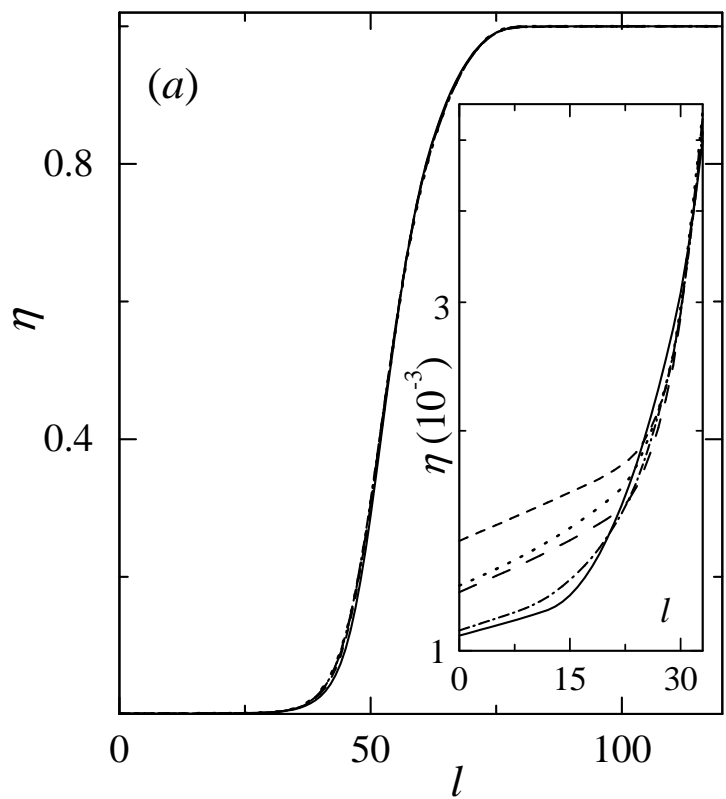
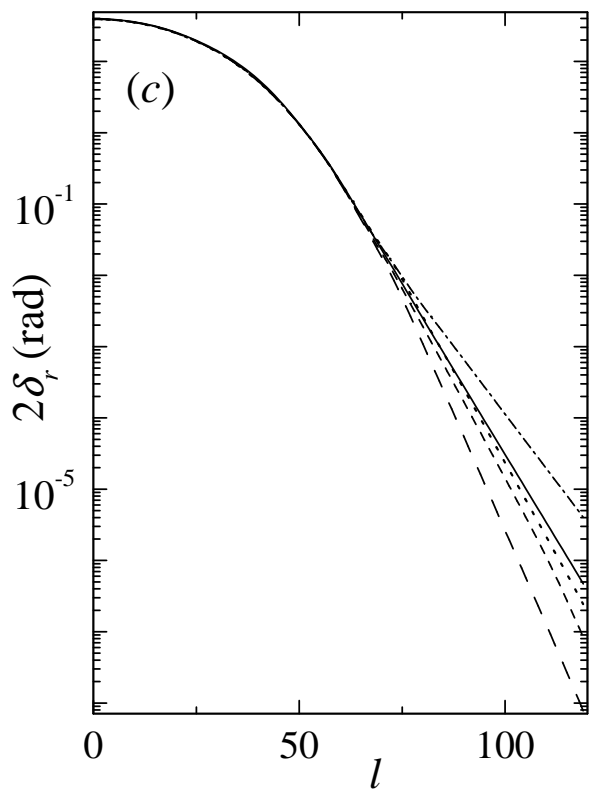
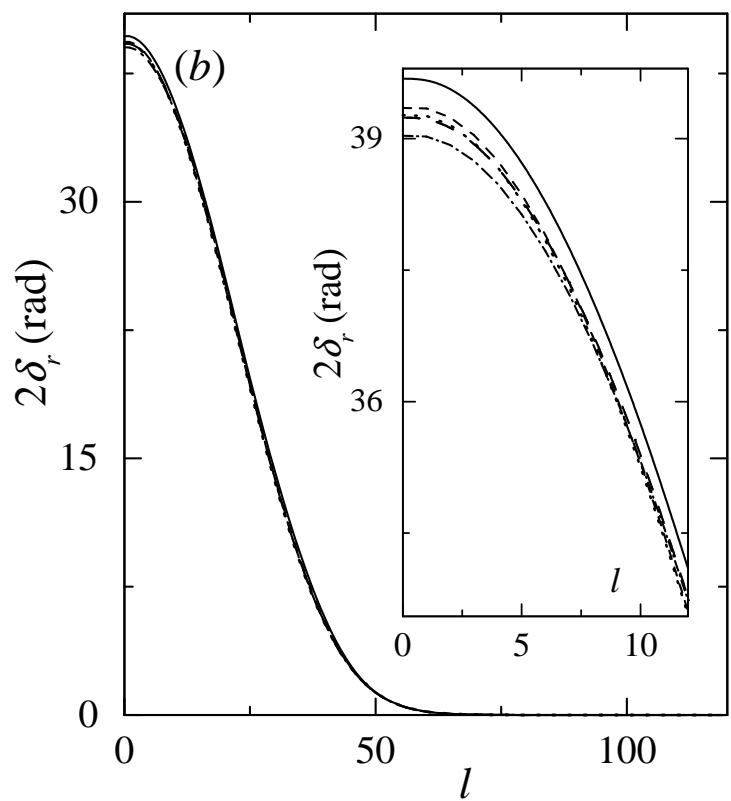
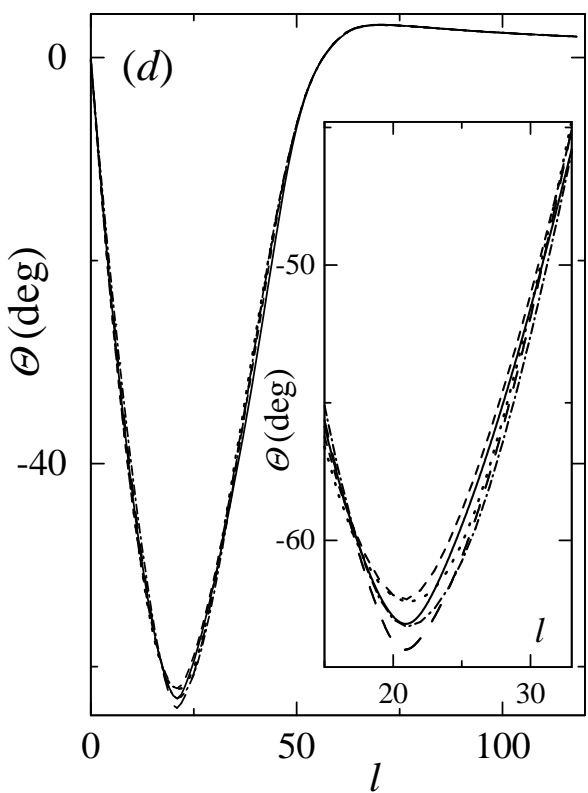
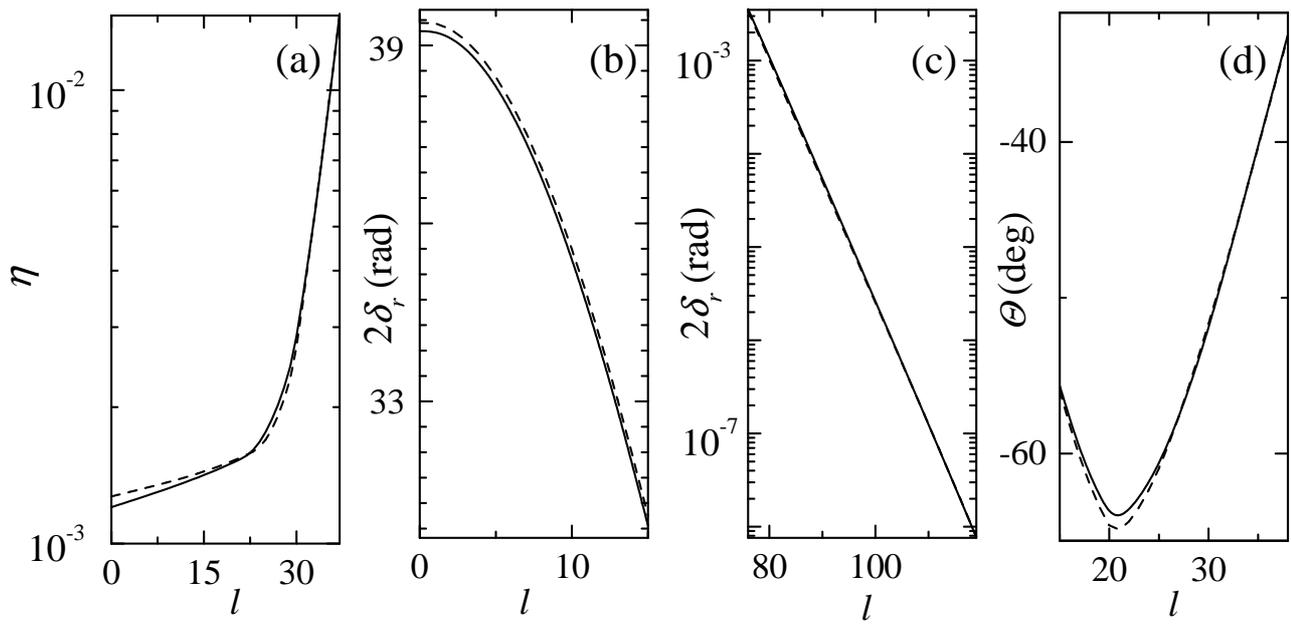
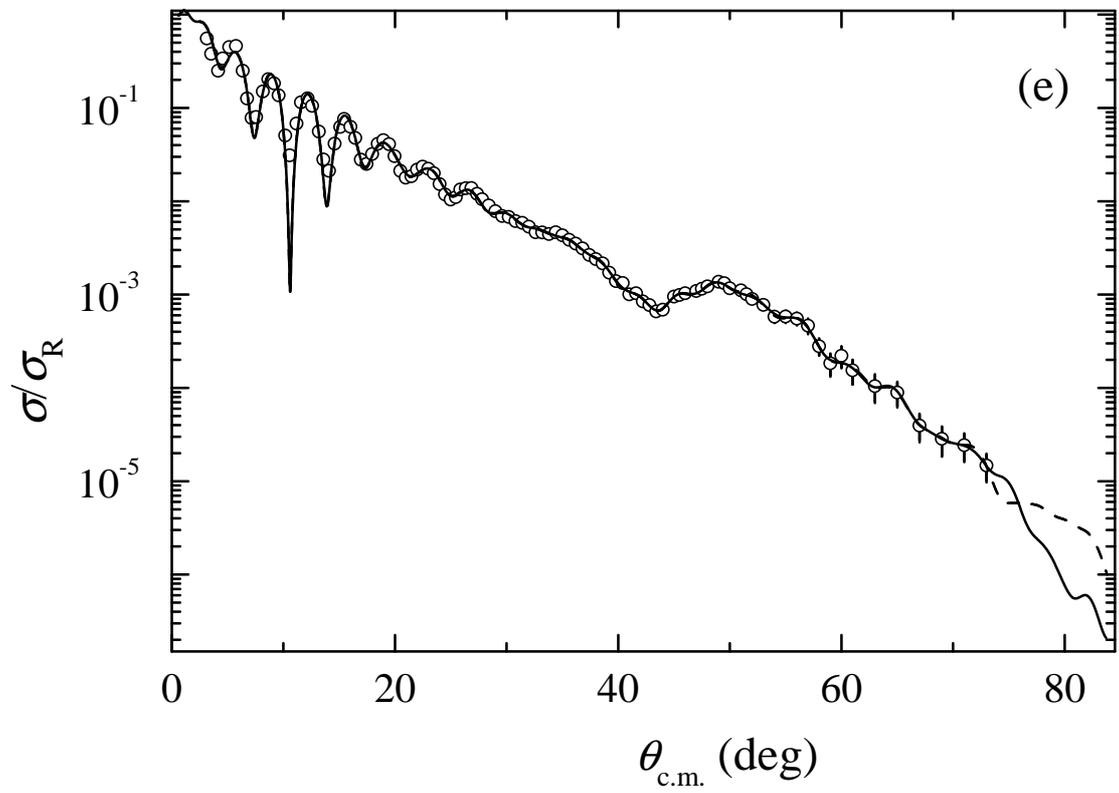
\end{document}